# Isolation of $^{236}$U and $^{239,240}$Pu from seawater samples and its determination by Accelerator Mass Spectrometry

*Mercedes López-Lora[1,2\*], Elena Chamizo[1], María Villa-Alfageme[3], Santiago Hurtado-Bermúdez[4], Núria Casacuberta[5,6], Manuel García-León[7]*

[1] Centro Nacional de Aceleradores (Universidad de Sevilla, Consejo Superior de Investigaciones Científicas, Junta de Andalucía), Thomas Alva Edison 7, 41092 Sevilla, Spain
[2] Departamento de Física Aplicada I, Escuela Politécnica Superior, Universidad de Sevilla, Virgen de África 7, 41011 Sevilla, Spain
[3] Dpto. de Física Aplicada II, Universidad de Sevilla, Av. Reina Mercedes 4A, 41012 Sevilla, Spain
[4] Servicio de Radioisótopos, Centro de Investigación, Tecnología e Innovación, Universidad de Sevilla, Av. Reina Mercedes 4B, 41012 Sevilla, Spain
[5] Laboratory of Ion Beam Physics, ETH Zürich, Otto-Stern-Weg 5, CH-8093 Zürich, Switzerland
[6] Environmental Physics, Universitätstrasse 16. CH-8092 Zürich, Switzerland
[7] Dpto. de Física Atómica Molecular y Nuclear, Universidad de Sevilla, Reina Mercedes s/n, 41012 Sevilla, Spain

\*Corresponding author:
E-mail address: mlopezlora@us.es

## Abstract

In this work we present and evaluate a radiochemical procedure optimised for the analysis of $^{236}$U and $^{239,240}$Pu in seawater samples by Accelerator Mass Spectrometry (AMS). The method is based on Fe(OH)$_3$ co-precipitation of actinides and uses TEVA® and UTEVA® extraction chromatography resins in a simplified way for the final U and Pu purification. In order to improve the performance of the method, the radiochemical yields are analysed in 1 to 10 L seawater volumes using alpha spectrometry (AS) and Inductively Coupled Plasma Mass Spectrometry (ICP-MS). Robust 80% plutonium recoveries are obtained; however, it is found that Fe(III) concentration in the precipitation solution and sample volume are the two critical and correlated parameters influencing the initial uranium extraction through Fe(OH)$_3$ co-precipitation. Therefore, we propose an expression that optimises the sample volume and Fe(III) amounts according to both the $^{236}$U and $^{239,240}$Pu concentrations in the samples and the performance parameters of the AMS facility. The method is validated for the current setup of the 1 MV AMS system (CNA, Sevilla, Spain), where He gas is used as a stripper, by analysing a set of intercomparison seawater samples, together with the Laboratory of Ion Beam Physics (ETH, Zürich, Switzerland).

**Keywords:** Accelerator Mass Spectrometry; 236U; Plutonium; Seawater; Preconcentration; Ion-exchange chromatography

# 1. Introduction

The naturally occurring $^{238}$U ($T_{1/2}$ =4.51·10$^9$ y), $^{235}$U ($T_{1/2}$ =7.038·10$^8$ y) and $^{234}$U ($T_{1/2}$ =2.455·10$^5$ y) have been extensively used in oceanographic studies, initially with the study of their distribution in the ocean and expanding to studies of their daughter isotopes as tracers of past and present marine processes [1]. Due to its conservative behaviour in open ocean the residence time of uranium in the marine environment as carbonate ion complex UO$_2$(CO$_3$)$_3$$^4$, is of about 400 ky [2]. Typical $^{238}$U concentrations in seawater are 3.5 ppb (3.5 µg L$^{-1}$ or 3 mBq L$^{-1}$) allowing for the use of conventional radiometric and mass spectrometry techniques for its analysis [3]. Recently, several studies have been focused on the determination of $^{236}$U ($T_{1/2}$ =2.342·10$^7$ y) that is essentially anthropogenic and it has been present in seawater since the onset of the nuclear age in 1945 [4-8]. Unlike the above mentioned isotopes, the measurement of $^{236}$U has been elusive for many years [9] because $^{236}$U concentrations in the environment, and particularly in the open ocean, are usually extremely low. For instance, surface seawaters from the Northern Hemisphere have mass concentrations at the level of fg L$^{-1}$ (i.e. nBq L$^{-1}$) [4, 5]; despite being strongly influenced by local and global fallout (i.e. aerosols released during the period of nuclear atmospheric testing, from 1945 to 1980). These $^{236}$U concentrations cannot be assessed by radiometric counting techniques with reasonable sample volumes. Therefore, the study of $^{236}$U in environmental samples and its use as an oceanographic tracer has only recently become possible as a result of the high sensitivity reached by modern Mass Spectrometry systems (MS). $^{236}$U/$^{238}$U atomic ratios in seawater are very low (from 10$^{-12}$ to 10$^{-9}$ [4, 5, 10]) and the highly abundant $^{235}$U and $^{238}$U may interfere to $^{236}$U analysis, due to their very similar trajectories in the different cinematic filters and to the formation of molecular isobars (e.g. $^{235}$U$^1$H). However, these two problems can now be overcome with Accelerator Mass Spectrometry (AMS). This technique has the lowest $^{236}$U/$^{238}$U background atomic ratios mainly by means of an optimized design of the mass spectrometer and of the destruction of the molecules in the so-called stripping process, which occurs in the terminal of an electrostatic tandem accelerator. Compared to other MS techniques such as TIMS (*Thermal Ionization Mass Spectrometry*) and HR-ICP-MS (*High Resolution Inductively Couple Plasma Mass Spectrometry*), conventional AMS systems (i.e. accelerator terminal voltages of 3 MV and above) can reach ratios down to 10$^{-14}$ – 10$^{-12}$, which are a few orders of magnitude better than TIMS and HR-ICP-MS (Table 1). In the last years, compact AMS systems, working at accelerator terminal voltages of 1 MV and below, have shown their potential to measure $^{236}$U at environmental levels [11, 12], reaching sensitivities at the 10$^{-12}$ for optimised designs [12].



Different from U, Pu is considered almost exclusively anthropogenic. The most abundant Pu isotopes are the fissile $^{239}$Pu(T$_{1/2}$ =24110 y) and $^{240}$Pu(T$_{1/2}$ =6564 y), which are produced by the $^{239}$Pu neutron activation in nuclear reactions. Plutonium has a very different geochemical behaviour than uranium; it is a particle-reactive element that is exported from the surface waters with sinking particles, and is re-dissolved in deep waters with the remineralisation of the particles. The chemistry of plutonium in seawater is complex because different valences plutonium isotopes coexist and constitute inorganic and organic ligands. Similar to $^{236}$U, Pu has been also introduced to the oceans by atmospheric weapon tests and the releases from nuclear facilities and most of this Pu is still present in the water column [13]. Another similarity with $^{236}$U is that plutonium concentrations in seawater are extremely low; e.g. $^{239+240}$Pu concentrations in surface samples from the Indian Ocean and South Atlantic Ocean are at the level of µBq L$^{-1}$ [14] and deep-water columns from the Arctic Ocean have $^{239+240}$Pu concentrations ranging 1-20 µBq L$^{-1}$ [15]. Therefore, high sensitivity techniques are required for its determination [16-19]. Alpha spectrometry (AS) was traditionally used for plutonium measurement in seawater samples [16, 19, 20]. The disadvantages in this case are: (i) large sample volumes, ~100 L, and long counting times due to a high limit of detection (usually 0.1 mBq) [21, 22] are needed and (ii) it is not possible to discriminate $^{239}$Pu and $^{240}$Pu isotopes in environmental samples, due their very similar alpha-energy emissions (5.1566 MeV and 5.1682 MeV, ΔE=11.6 keV). These drawbacks are overcome by MS techniques, However, the detection limit reached by HR-ICP-MS, which is the most used technique in this field [23], is highly limited by the presence of uranium hydrides, $^{238}$UH$^{+}$ and $^{238}$UH$^{2+}$, which cannot be resolved from $^{239}$Pu$^{+}$ and $^{240}$Pu$^{+}$. In this context, AMS stands out for setting low detection limits regardless of the matrix components of the sample, because it is able to eliminate molecular isobars in the so-called stripping process [24].

There is a great variety of published methods for plutonium and uranium separation from different sample matrixes using different techniques. As far as seawater studies are concerned, the situation is different for those two elements. Due to the extremely low plutonium concentration involved, most of the reported procedures deal with large sample volumes (i.e. at the 100 L range) [25]. However, much lower volumes have been traditionally processed in the case of the naturally occurring uranium isotopes given their concentration in seawater [26]. On the other hand, $^{236}$U concentration in seawater is significantly lower, thus in the last years an important effort have been devoted to the implementation and optimization of methods for the uranium extraction in order to analyse $^{236}$U by AMS [4, 8, 27, 28]. Most of the published procedures have in common the initial pre-concentration from the bulk sample by Fe(OH)$_3$ precipitation. In a recent work [28] using 10 L samples, it is shown that uranium recovery in that pre-concentration step can be optimised by adjusting the amount



of iron and applying $N_2$ bubbling for the subsequent sample degasing. However, more simple methods are required for oceanographic expeditions where the pre-concentration step is usually developed on research vessels, $N_2$ bubbling is not always possible and it is key to minimize the processing time and to maximize the number of samples that can be collected and processed on board. On the other hand, given the high value and uniqueness of the samples involved in oceanographic expeditions many efforts have been dedicated to develop sequential methods for the extraction of different elements from the same sample [29-31].

In this work, we present a radiochemical method to separate sequentially U and Pu fractions for $^{236}$U and $^{239,240}$Pu AMS determinations, and study the corresponding recovery yields by AS and ICP-MS. The method is based on the commonly used co-precipitation of actinides with $Fe(OH)_3$, and uses TEVA® and UTEVA® resins in a very simplified way for the final U and Pu purifications. The evaluation of the method is focused on the analysis of the $Fe(OH)_3$ pre-concentration step for the typical seawater volumes involved in an AMS analysis, from 1 to 10 L. In order to further understand the role of the iron on the key co-precipitation stage, a comprehensive study of the correlation between the amount of added iron and the seawater sample volume was performed. The method for AMS measurements was validated through the analysis of a set of intercomparison seawater samples, together with the Laboratory of Ion Beam Physics (ETH, Zürich, Switzerland). Finally, the figures of merit for an efficient $^{236}$U and $^{239,240}$Pu AMS analysis at an AMS facility were evaluated and optimised for the Centro Nacional de Aceleradores (CNA, Sevilla, Spain) as a case study.

## 2. Experimental

**2.1 Samples**

For the evaluation of the radiochemical method described in the following sections, seawater samples collected in the Isla Canela's beach (Huelva, Spain) were used. About 200 L of seawater were taken during two sampling campaigns in 2014 and 2015. Samples were filtered using a 60-68 µm filter to remove any traces of sand or seaweed. Aliquots ranging from 1 to 10 L were considered for the different tests. These are the standard volumes that one can get during an oceanographic expedition where samples are taken from a normal CTD-rosette (12 L Niskin bottles). For the validation of the method proposed in this paper, a set of intercomparison seawater samples provided by the ETH Zürich were processed and measured on the 1MV AMS system at CNA.



## 2.2 Spikes, materials and reagents

$^{238}U$ is an abundant natural isotope and any trace of $^{238}U$ in reagents and laboratory equipment would underestimate the measured $^{236}U/^{238}U$ atom ratios, which are usually the result of an AMS measurement. Also the presence of $^{238}U$ might produce background effects on $^{239}Pu$. To minimize the uranium contamination during the sample processing, *Suprapure* grade reagents and MQ water (i.e. Milli-Q® water, an ultra-pure distilled water) are used. Besides, glassware is avoided when possible. For the $Fe(OH)_3$ co-precipitation step, a $Fe(III)$ solution provided by *High Purity Standards* (HPS, England) was used, with a certified $^{238}U$ concentration below the 0.5 ppb level. According to our experience, no additional $^{236}U$ contamination can be attributed to this iron solution.

Ion chromatography separations were performed in a vacuum box and using TEVA® and UTEVA® resins (Eichrom Industries, Inc.). 2 mL cartridges of 100-150 µm particle size material were used for the experiments [32, 33].

Two different spikes were used for the AMS measurements, $^{233}U$ and $^{242}Pu$. $^{233}U$ was provided by the IAEA Environment Laboratories in Monaco (unknown supplier) and contains $^{236}U$ at the $10^{-3}$ level (i.e. $^{236}U/^{233}U$ atom ratio). $^{242}Pu$ was purchased from the National Physical Laboratory (NPL, England) and presents $^{239,240}Pu/^{242}Pu$ and $^{236}U/^{242}Pu$ atom ratios at the $10^{-6}$ level [15].

$^{232}U$ ($T_{1/2}$=68.9 y) was used as spike for AS determinations and was purchased from the *Centro de Investigaciones Energéticas, Medioambientales y Tecnológicas* (CIEMAT, Spain).

## 2.3 Radiochemical method for $^{236}U$ and $^{239,240}Pu$ isotopes determination by AMS

The different steps of the radiochemical method used at the CNA are schematised in Fig. 1 and explained next.

### 2.3.1. Sample pre-treatment

About 3 pg of $^{233}U$ ($T_{1/2}$= 1.59·$10^5$ y) and $^{242}Pu$ ($T_{1/2}$=3.73·$10^5$ y) spikes were initially added to the sample to trace the radiochemical procedure and finally calculate the concentrations of $^{236}U$ and $^{239,240}Pu$ (isotope-dilution method). Samples were acidified with 1 mL of $HNO_3$ (65%) per litre of sample, mixed and let equilibrate for at least one hour to ensure homogenisation. $Fe(III)$ was then added to the samples and subsequently vigorously shaken to guarantee homogeneity (see discussion in section 3.3). Uranium and plutonium were extracted from the bulk of the sample (i.e. from 1 to 10 L of seawater) by $Fe(OH)_3$ coprecipitation, raising the pH to about 8.5 with 45% $NH_3$. After a complete precipitation and settling of the $Fe(OH)_3$ particles, the supernatant was discarded and the resulting $Fe(OH)_3$ precipitate transferred to 125 mL plastic bottles and centrifuged at 3500 rpm



during 5 minutes. A final washing of the precipitate with MQ water was applied to clean the precipitate from salts.

*2.3.2 Separation of the Pu and U fractions with TEVA® and UTEVA® resins*

Fe(OH)$_3$ precipitate was transferred to a Teflon® beaker, dried and, finally, dissolved with 20 mL of 8M HNO$_3$. At these conditions, the acid dependency curves for the TEVA® resins showed a high uptake for Pu(IV), Np(IV) and Th(IV), whereas the UTEVA® showed the maximum sorption for U(VI) [33]. Therefore, TEVA® and UTEVA® resins were arranged in a vacuum box in tandem, as shown in Fig. 2. Sample was loaded into both resins where Pu(IV) is initially adsorbed onto the TEVA® resin and U(VI) does onto the UTEVA® resin. In order to have Pu(IV) oxidation state, 0.2 mL of 3M NaNO$_2$ were added to each sample previous to this step.

When both radionuclides were in the suitable oxidation state, Pu and U separation was performed as follows. First, TEVA® and UTEVA® resins were washed with 10 mL of MQ water and conditioned to nitric acid media by adding 20 mL of 8M HNO$_3$. Afterwards, the sample was loaded onto the stacked columns: plutonium was sorbed to the TEVA® resin and uranium to the UTEVA® resin. Finally, both resins were separated in order to elute plutonium from TEVA®, by adding 15 mL of a solution 0.1M HCl/0.01M HF and 10 mL of MQ water, and uranium from UTEVA®, by adding 15 mL of 0.1M HNO$_3$ and 10 mL of MQ water. In every step, the residues were stored in order to check the presence of uranium and plutonium if necessary.

**2.4 Evaluation of the U and Pu radiochemical yields using AS**

In order to optimised the proposed method, in a first approach, the U and Pu recoveries were quantified by AS. The aim of this first experiment was to study the possible influence of the salinity and/or the total seawater volume in the final U and Pu recoveries. To this end, a set of samples were prepared using different volumes of seawater or a mixture of seawater and MQ water (Table 2). $^{233}$U and $^{242}$Pu were used as initial spikes (i.e. 70 mBq of $^{232}$U and 15 mBq of $^{242}$Pu were added in the pre-treatment step). The AS sources were prepared from the U and Pu final solutions following the electrodeposition method described in [34]. The recoveries of this electrodeposition step were quantified independently and excluded from the final U and Pu radiochemical yields.

In a first set of samples (from MLL-1 to MLL-6) both U and Pu recoveries were studied (Table 2). The electrodeposition yield was obtained from an additional set of samples that were exclusively electrodeposited (i.e. blank solutions spiked with $^{232}$U and $^{242}$Pu). The obtained electrodeposition yields were (96±4)% for uranium and (87±4)% for plutonium.



In a second exercise (samples from MLL-10 to MLL-16), only U recoveries were quantified because of the coherent results obtained from the previous exercise for Pu. These samples were double-spiked, first with $^{232}$U to quantify the overall uranium recovery and additionally, after the extraction chromatography step, with $^{233}$U to exclusively control the electrodeposition yield.

The AS determinations were performed at the Radioisotopes Service of the University of Sevilla (CITIUS, *Centro de Investigación, Técnología e Innovación de la Universidad de Sevilla*). This facility is equipped with an array of 24 chambers provided with passivated implanted planar silicon detector (PIPS, 18 keV of nominal resolution) [35].

**2.5 Evaluation of uranium co-precipitation with Fe(OH)$_3$ using ICPMS**

In a second experiment, ICP-MS was used to evaluate uranium radiochemical yields during the Fe(OH)$_3$ co-precipitation process by analysing the naturally present $^{238}$U in the samples. To this end, the $^{238}$U concentrations in the original samples and in the resulting supernatants were measured by ICP-MS and compared (Fig.1).

In order to check the influence of the Fe concentration in the sample, different amounts of Fe(III) were added to 1 L aliquots. On the other hand, the effect of the total amount of seawater was also studied in a second experiment using samples with different volumes and variable amounts of Fe(III) to maintain Fe concentration constant. Finally, to study the effect of acidification time (i.e. time lapse between acidification and pre-concentration), 4 seawater samples of 10 L were processed as follows: two samples were acidified 24 hours prior to Fe(OH)$_3$ co-precipitation and the other two samples 10 days.

An Agilent 7500c ICP-QMS provided with an Octopole Reaction System (ORS) at the Radioisotopes Service of the University of Sevilla was used for these determinations [36].

**2.6 AMS Measurement**

The final aim of the proposed method was to analyse $^{236}$U and $^{239,240}$Pu by AMS. To finally place the samples into AMS cathodes (i.e. the final sample adapted to the optimal matrix for the actinides extraction in the ion-source), the final uranium and plutonium solutions were co-precipitated as Fe(OH)3 after adding 1 mg of Fe(III) to the solution. Precipitates were transferred to quartz crucibles, dried and oxidized at 650°C. Finally, samples were mixed with 3 mg of Nb powder and pressed into aluminium cathodes. [15].

AMS determinations were performed on the 1 MV AMS system at the CNA. The setup of the facility for the measurement of Pu isotopes was discussed in [37], and the parameters of the $^{236}$U measurement using Ar gas as stripper were reported in [11]. Briefly, plutonium or uranium isotopes



are extracted from the Cs$^+$ sputtering ion-source as negative oxide ions (i.e. PuO$^-$, UO$^-$). These anions are analysed by a first 90º sector magnet and directed to the terminal of the tandem accelerator working at about 650 kV, which contains the stripper gas, where they are dissociated and stripped to 3+ charge state in the so-called stripping process (i.e. Pu$^{3+}$, U$^{3+}$). These ions are analysed by a second 90º sector magnet and by a 120º electrostatic deflector. Finally, the different isotopes ($^{233,236}$U$^{3+}$, $^{239,240,242}$Pu$^{3+}$) are counted from the total energy signal provided by a gas ionization chamber. In the case of uranium measurements, $^{238}$U$^{3+}$ is measured as a beam-current in an off-axis Faraday cup placed at the exit of the second sector magnet. Recently, the stripper gas of the facility has been changed from Ar to He, following the promising results obtained in other AMS facilities (i.e. higher stripping yields specially for the heaviest radionuclides [38]). Besides, a miniaturized gas detector designed and manufactured by ETH Zürich, has been installed. Details about the current status of the 1 MV CNA AMS system are given in [39] and [40].

## 3. Results and discussion

### 3.1 Evaluation of the U and Pu radiochemical yields

In order to further study the proposed method, we focused on the possible influence of the salinity and sample volume in the U and Pu recoveries by analysing 14 samples (Table 2) by AS as it was described in the section 2.4.

The obtained Pu radiochemical yields were satisfactory for most of the samples and range from 70% to 88% (the low yield for MLL-2, i.e. 51±6%, might be caused by a problem in the electrodeposition step; furthermore, this result was excluded from the discussion). In contrast, the obtained U recoveries were not considered acceptable (i.e. over 50%) in all the cases. (Table 2)

Fig. 3 shows the dependence of the obtained radiochemical yields for uranium and plutonium with the volume of the sample. In samples composed entirely or partly by seawater, uranium yields clearly decreased with seawater volume. Furthermore, similar uranium recoveries were obtained for samples MLL-3 (i.e. 5 L of MQ water plus 5 L of seawater) and MLL-5 (i.e. 5 L of seawater), which indicate that: (i) the decrease of the yield does not depend on the salinity of the sample, and (ii) the total sample volume is not influencing the recovery and it is most likely influenced by the volume of seawater. Additionally, the results obtained for the two MQ water samples mixed with 120 g or 270 g of NaCl (samples MLL-15 and MLL-16, respectively) indicate that the total amount of NaCl, which is one of the most relevant components of seawater, does not influence the chemical yield.

It is worth mentioning that similar results were obtained from the study of the $^{238}$U, which is naturally present in seawater, throughout the procedure. In Table 3, the obtained radiochemical



yields by AS using the artificially added $^{232}$U spike and the natural $^{238}$U are displayed. The fact that there is no isotopic fractionation during the radiochemical separation is very important in relation to the measurement of anthropogenic $^{236}$U; as it might be present in the sample in different physical-chemical forms than the primordial $^{238}$U, due to their very different origins and residence times in the oceans.

To get more information about the behaviour of the uranium fraction during the different steps of the proposed procedure, the presence of natural $^{238}$U in the different residues produced in every step was characterized by ICP-MS. No traces of $^{238}$U were found in the wastes produced during the extraction chromatography step, however, significant concentrations of that radionuclide were found in the supernatant produced in the Fe(OH)$_3$ pre-concentration step. Indeed, those concentrations were correlated with the total seawater volume similarly as the total uranium recoveries obtained by AS, pointing out that this initial step accounts for most of the uranium losses during the radiochemical process. This agrees with other studies [28] and therefore, this critical step was studied in detail and it is discussed next.

### 3.2 Optimization of the uranium co-precipitation with Fe(OH)$_3$

Two factors can play a potential critical role during the Fe(OH)$_3$ co-precipitation process (i) the concentration of Fe(III) in the sample (i.e. the total amount of Fe(III) added before the co-precipitation step for each litter of sample), and (ii) the total volume of processed seawater. The first one was studied by adding different Fe(III) amounts to 1 L samples and quantifying the recovery of $^{238}$U during the Fe(OH)$_3$ co-precipitation by ICPMS. The obtained results are plotted in Fig.4

Two relevant conclusions are reached from these results: the uranium recovery from seawater, $Y$, (i) depends strongly on the Fe(III) concentration in the sample, $c$, and (ii) increases with Fe(III) concentration following the asymptotic trend given:

$$Y = Y_{max} \frac{c}{k+c} \qquad (1)$$

where $Y_{max}$ represents the maximum recovery that it could be obtained and $k$ the Fe concentration needed to achieve a half-maximum yield.

For 1 L sample, the dataset were fitted to the asymptotic curve as, $Y_{max} = (87.8 \pm 3.8)$ % and $k = (11.6 \pm 2.1)$ mg L$^{-1}$. The maximum uranium recovery would be approximately 80% and could be reached for Fe(III) concentrations in the sample above 80 mg L$^{-1}$. Higher concentrations of iron do



not significantly increase uranium recoveries and, on the contrary, according to our experience, might alter the performance of the resins for seawater volumes higher than 5 L.

The effect of the volume of seawater was studied by analysing samples with the same Fe(III) concentration but processing different volumes (Fig. 5). At a first glance, the results show that when the Fe(III) concentration increases, uranium recoveries are improved. Therefore, that means that the low recoveries obtained in section 3.1 when processing 10 L aliquots might be partly caused because not enough iron was added during the co-precipitation process (i.e. 20 mg $L^{-1}$). However, these results also show that total seawater volume is another important factor to consider. This is especially clear if we compare the results from the 20 and 150 mg $L^{-1}$ of Fe(III) in different seawater volumes. In the case of 150 mg $L^{-1}$, uranium radiochemical yield decreased from 80% to 70% when seawater volume was increased from 1 to 10 L. However, when 20 mg $L^{-1}$ of Fe(III) was added, recoveries decreased from 50% to 25%. That means that the reduction is not constant and is more significant for the lower Fe(III) concentrations. Therefore, the influence of the iron concentration on the uranium recovery is correlated with the sample volume.

These results evidence the different biogeochemical behaviour in seawater of the two elements of interest in our study. In contrast with uranium results, the recovery of plutonium during the $Fe(OH)_3$ co-precipitation process does not seem to be either affected by the Fe(III) concentration or seawater volume, within the range of values considered, probably due to the strong particle-reactive nature of this element. The influence of the volume in the uranium recoveries is not easily well understood and it might be related to its high solubility in this medium, which is strongly related to the presence of carbonates [41, 42]. Uranium exists as uranyl ions in seawater forming carbonate complexes, $UO_2(CO_3)_3^{-4}$, which are relatively stronger than many ligands and might remain in solution in the co-precipitation process. Therefore, in order to precipitate uranium, these uranyl carbonate complexes have to be completely dissociated.

In the pretreatment step, before the $Fe(OH)_3$ precipitation stage, samples are acidified with $HNO_3$ (i.e. pH=2) during 1 hour in order to dissociate uranyl carbonate complexes into uranyl ion and carbonic acid [43]. Thus, the free uranyl ions are able to associate to $Fe(OH)_3$ colloids during the coprecipitation process. In Table 4, uranium recoveries for different acidification times (i.e. 24 hours and 10 days) are reported. No significant differences are observed in the recoveries of these samples and the previous ones (Fig. 5). Therefore, we assume that the uranyl carbonate complexes were almost completely dissociated before $Fe(OH)_3$ co-precipitation so they cannot be the main reason of the decrease in the recovery.

Note that the influence of the iron on the performance of the TEVA and UTEVA resins and, therefore, on the overall radiochemical Pu and U yields, has not been further studied. For this reason,



in summary, we do not recommend to use iron in excess during the co-precipitation process. We find necessary to reach a compromise of the optimal amount of Fe in each case. For example, in the most critical case (i.e. 10 L seawater aliquots), Fe(III) concentrations of 80 mg L$^{-1}$ are recommended. In these conditions, there will be a total amount of 800 mg of Fe in the sample, and this amount of Fe could be completely removed with TEVA® and UTEVA® resins according to our experience. In these conditions, plutonium radiochemical yield would be around 80% and uranium yield close to 60%.

### 3.3 Validation of the method for AMS

The final aim of the proposed method was to measure $^{236}$U and Pu isotopes by AMS from seawater. In order to validate the method, a set of intercomparison samples were measured on the 1 MV CNA AMS system. The obtained results are shown in Table 5. The samples were already used in a previous intercomparison exercise between the CNA AMS system and the 500kV Tandy AMS system at ETH (Zürich). Those samples were initially measured following an alternative chemical procedure and using a different setup of the CNA AMS system [15].

In this new exercise, only 3-5 L of seawater were processed (versus 5-9 L that were used in the previous intercomparison [15]). In general, there is a fine agreement between previous and the present results and between ETH and CNA. Besides, a better statistical precision, especially for plutonium measurements, was obtained this time in spite of the smaller processed seawater volumes. In these AMS measurements, the overall counting efficiency was improved as a result of the implementation of the He gas as stripper (i.e. transmission of actinides through the tandem accelerator was increased from 11% to 38% [40]).

The results from Table 5 validate the radiochemical method and the implemented AMS setup at CNA [40]. Note that in Table 6, $^{239,240}$Pu activity concentrations are only shown for CNA measurements, since this information was not provided by the ETH, as no plutonium spike was added to those samples.

### 3.4 Optimization of the sample volume for AMS measurements at Centro Nacional de Aceleradores (CNA)

In order to individually optimize the seawater volume required for a reliable $^{236}$U and $^{239,240}$Pu AMS determination, two issues must be analysed: (i) losses during the radiochemical process, and discussed in the previous sections, are expected and (ii) the performance of the specific AMS measurement system must be considered.



We aim to quantify both features into a single expression. For a volume *V* of sea water, we propose to estimate the minimum concentration able to be quantified by AMS ($c_{min}$) using the following expression,

$$c_{min} = \frac{D\, t}{\varepsilon\, RY\, V} \quad (2)$$

Where *RY* is the radiochemical yield of the method, *D* is the background corrected detector signal (counts per second), *t* is the measurement time and ε is the detection efficiency of the AMS system for a specific element. Note that, in the case of uranium determinations, the *RY* depends on the samples volume and can be estimated in advance, as it was discussed in the previous sections.

This expression is universal for any AMS facility, and only the detection efficiency term should be adapted. This term, which is the ratio between the atoms that are actually detected and the ones present in the cathode (i.e. the final AMS sample), can be quantified by analysing a series of samples artificially spiked with well-known amounts of different radionuclides (i.e. $^{233}$U and $^{242}$Pu in our case).

For the AMS system at CNA, typical count rates are 100 cps and 200 cps per pg of $^{233}$U and $^{242}$Pu in the cathode, respectively. This means that, for a typical measurement time of 30 minutes, we got an overall detection efficiency of $7\cdot 10^{-5}$ and $1.5\cdot 10^{-4}$ for uranium and plutonium isotopes, respectively.

In order to obtain the minimum concentration of any of the radionuclides of interest (i.e. $^{239,240}$Pu and $^{236}$U) that can be evaluated at CNA, 100 detected atoms, which means 100 net counts in the ionization chamber after the background correction, were considered (i.e. 10% statistical precision). The background counts are estimated using the count rates produced by instrumental blanks, which are cathodes with the same matrix as the samples of interest but free of those radionuclides (i.e. iron oxide mixed with Nb in the same mass ratio as the sample of interest). In our case, a typical count rate of 0.015 cps is obtained in the ionization chamber for those blank samples. Therefore, for a 30 min measurement, typically 27 counts are detected as background events.

From the equation (2), we can estimate the minimum concentration of $^{236}$U and Pu isotopes that can be assessed as a function of the sample volume (Fig. 6). For these estimations, the influence of the volume processed in the uranium radiochemical yield (*RY*) has been also considered.

In the case of Pu measurements, as recoveries are high and constant within this range of studied volumes, the main limitation is imposed by the efficiency of the AMS system. Concentrations below $10^5$ atoms L$^{-1}$ can be assessed for up to 10 L seawater samples. For example, that means that, according to the reported $^{239}$Pu concentrations in [14] (between $6\cdot 10^5$ atoms L$^{-1}$ to $1\cdot 10^6$ atoms L$^{-1}$), 2 L of sample would be enough to analyse this radionuclide in surface seawater samples from the Indian Ocean [14]. However, to afford the measurement of both $^{239}$Pu and $^{240}$Pu, the minimum volume is determined by the less abundant $^{240}$Pu. For example, $^{240}$Pu concentrations at the order of $10^5$-$10^6$ atoms L$^{-1}$ have been reported in the Arctic Ocean [15]. Hence, volumes up to 10 L are large enough to assess $^{240}$Pu concentrations at these levels.

For $^{236}$U measurements two different cases are plotted in Fig. 6, depending on the amount of Fe(III) used in the co-precipitation step. The influence of the iron concentration is especially important when $^{236}$U concentrations below $10^6$ atoms L$^{-1}$ are expected. However, the main limitation comes from the abundance sensitivity provided by the AMS system (i.e. minimum $^{236}$U/$^{235}$U atom ratio can be actually determined). $^{235}$U, which is naturally present in the sample, can mimic the trajectory of $^{236}$U in the different cinematic filters due to its very similar masses, producing spurious events in the detector. If the resulting $^{236}$U/$^{235}$U atom ratio is multiplied by the natural abundance of $^{235}$U (i.e. 0.7204%), the limiting $^{236}$U/$^{238}$U atom ratio is obtained. As it was stated before, this is the figure of merit of an AMS system that has to be taken into account when the measurement of $^{236}$U in environmental samples is considered. In our facility, ratios below $9 \cdot 10^{-11}$ cannot be measured at the 1 MV CNA setup [40], regardless of the volume of the processed sample. Fig. 7 presents the minimum $^{236}$U/$^{238}$U atomic ratio as a function of the sample volume considering the obtained experimental results, and including the $^{236}$U/$^{238}$U limit at CNA. According to Fig. 7, we conclude that in our case a 4-litre seawater sample is enough to overcome the limitation imposed by the $^{236}$U/$^{238}$U atom ratio.

For what the oceanographic applications are concerned, surface seawater samples are measurable on the CNA AMS system (i.e. $^{236}$U/$^{238}$U atom ratios above the $10^{-10}$ level); $^{236}$U/$^{238}$U ratios ranging from about $10^{-10}$ to about $7 \cdot 10^{-10}$ were reported for a seawater column sampled in the North Atlantic Ocean [7]; and very recently, $^{236}$U/$^{238}$U atom ratios above $10^{-9}$ were measured in the Northwestern Mediterranean Sea [44].

## 4. Conclusions

A radiochemical method for the sequential separation of uranium and plutonium from low-volume seawater samples has been established at CNA. Radiochemical yields of plutonium range from 70% to 88% in sample volumes from 1 to 10 litres, with no significant effect of the seawater volume or presence of salts in the Fe(OH)$_3$ precipitates. On the contrary, uranium yields depend on (i) the concentration of Fe(III) in the sample, and (ii) the initial volume of the seawater sample, during the first pre-concentration, which is the most critical step of the method. Uranium co-precipitation recoveries increase asymptotically with increasing Fe concentration and general uranium yields decrease with sample volume. Therefore, uranium yield can be optimised by adding higher amounts of Fe(III). However, using iron in excess (i.e. more than 800 mg) is not recommended because it could influence on the performance of the TEVA and UTEVA resins. In the most critical case, i.e. 10 L seawater samples, concentrations of 80 mg of Fe(III) per litre of sample are recommended and plutonium yields around 80% and uranium yields close to 60% are expected.

The proposed method is now optimised for the measurement of $^{236}$U and $^{239,240}$Pu by AMS in seawater volumes ranging from 1 to 10 L. These volumes are appropriate for a routinely oceanographic expedition using CTD-rosettes with 12L Niskin bottles. Pretreatment and coprecipitation steps could be also done on board. In the case of U and Pu measurements by AMS at CNA, the minimum $^{236}$U and Pu concentrations and $^{236}$U/$^{238}$U atomic ratios that can be quantified (i.e. 10% statistical precision) were studied as a function of the sample volume. Pu concentrations above $10^5$ atoms L$^{-1}$ can be assessed with volumes up to 10 litres. In the case of $^{236}$U, 4 L of sample are enough to overcome the $^{236}$U/$^{238}$U limit of 9·10$^{-11}$ for CNA AMS facility. The results of the intercomparison exercise validate the proposed method and demonstrate the potential of the compact AMS system at CNA for $^{236}$U and $^{239,240}$Pu determination in oceanographic samples.

**Acknowledgments**

This work has been financed from the project FIS2015-69673-P, funded by the Spanish Ministry of Economy. This work was partially funded by Fundación Cámara Sevilla through a Grant for Graduate Studies (M.L.L).

**Tables**

*Table 1 –Minimum $^{236}U/^{238}U$ atom ratios that can be evaluated by different MS techniques: TIMS (Thermal Ionization Mass Spectrometry), HR-ICP-MS (High Resolution Inductively Couple Plasma Mass Spectrometry) and AMS (Accelerator Mass Spectrometry).*

|  |  | $^{236}U/^{238}U$ **Sensibility** |
|---|---|---|
|  | **TIMS** | $2 \cdot 10^{-10}$ [45] |
|  | **HR-ICP-MS** | $1 \cdot 10^{-9}$ [46] |
| **AMS** | ANU (3.5 MV) - Australia | $1 \cdot 10^{-13}$ [47] |
|  | ETH-Tandy (500 kV) - Switzerland | $1 \cdot 10^{-14}$ [12] |
|  | CNA-SARA (1 MV) - Spain | $3 \cdot 10^{-11}$ [11] |

*Table 2 – Samples processed and measured by alpha spectrometry to study radiochemical yield of uranium and plutonium. Some samples were spiked with $^{232}U$ and $^{242}Pu$ tracers and others were only spiked with $^{232}U$. Electrodeposition blank samples were used to control electrodeposition process in samples from MLL-1 to MLL-6 (Blk-Ed); samples from MLL-10 to MLL-16 were spiked with $^{233}U$ in the last step to control the electrodeposition process directly.*

| Sample | Description | Spikes | | U Yield (%) | Pu Yield (%) |
|---|---|---|---|---|---|
|  |  | Total process | Electrodeposition |  |  |
| MLL-1 | 10 L MQ water | $^{232}U$, $^{242}Pu$ | Blk-Ed | 66 ± 4 | 86 ± 6 |
| MLL-2 | 10 L MQ water | $^{232}U$, $^{242}Pu$ | Blk-Ed | 69 ± 5 | 51 ± 6 |
| MLL-3 | 5 L MQ water + 5 L seawater | $^{232}U$, $^{242}Pu$ | Blk-Ed | 48 ± 4 | 72 ± 5 |
| MLL-4 | 7 L MQ water + 3 L seawater | $^{232}U$, $^{242}Pu$ | Blk-Ed | 74 ± 5 | 88 ± 6 |
| MLL-5 | 5 L seawater | $^{232}U$, $^{242}Pu$ | Blk-Ed | 48 ± 3 | 81 ± 6 |
| MLL-6 | 10 L seawater | $^{232}U$, $^{242}Pu$ | Blk-Ed | 11.1 ± 0.8 | 70 ± 5 |
| MLL-10 | 10 L seawater | $^{232}U$ | $^{233}U$ | 26 ± 4 | - |
| MLL-11 | 10 L MQ water + 10 L seawater | $^{232}U$ | $^{233}U$ | 31 ± 5 | - |
| MLL-12 | 10 L seawater | $^{232}U$ | $^{233}U$ | 12.1 ± 1.8 | - |
| MLL-13 | 10 L MQ water | $^{232}U$ | $^{233}U$ | 74 ± 11 | - |
| MLL-14 | 10 L MQ water | $^{232}U$ | $^{233}U$ | 68 ± 10 | - |
| MLL-15 | 10 L MQ water + 120 g NaCl | $^{232}U$ | $^{233}U$ | 77 ± 11 | - |
| MLL-16 | 10 L MQ water + 270 g NaCl | $^{232}U$ | $^{233}U$ | 76 ± 11 | - |

*Table 3 – Uranium radiochemical yield from natural $^{238}U$ and $^{232}U$ (tracer) obtained by AS. $^{238}U$ concentration, which was measured by ICP-MS, was $(4.0 \pm 0.3)$ µg $L^{-1}$ in these samples.*

| Sample | Description | U Yield (%) | |
|---|---|---|---|
|  |  | from $^{232}U$ | from $^{238}U$ |
| MLL-3 | 5 L MQ water + 5 L seawater | 48 ± 4 | 40 ± 4 |
| MLL-4 | 7 L MQ water + 3 L seawater | 74 ± 5 | 77 ± 8 |

| Sample | Volume | | |
|---|---|---|---|
| MLL-5 | 5 L seawater | 48 ± 3 | 45 ± 4 |
| MLL-6 | 10 L seawater | 11.1 ± 0.8 | 12.2 ± 1.2 |
| MLL-10 | 10 L seawater | 26 ± 4 | 20.1 ± 2.0 |
| MLL-11 | 10 L MQ water + 10 L seawater | 31 ± 5 | 28 ± 3 |
| MLL-12 | 10 L seawater | 12.1 ± 1.8 | 10.6 ± 1.1 |

*Table 4 – Uranium radiochemical yield in the Fe(OH)$_3$ coprecipitation step measured by ICP-MS. Before Fe(OH)$_3$ coprecipitation, different times of acidification were used.*

| Seawater volume (L) | Fe(III) concentration (mg L$^{-1}$) | Time of acidification | Uranium radiochemical yield (%) |
|---|---|---|---|
| 10 | 20 | 24 hours | 32 ± 6 |
| 10 | 20 | 24 hours | 20.2 ± 1.1 |
| 10 | 20 | 10 days | 23 ± 4 |
| 10 | 20 | 10 days | 29.1 ± 1.3 |

*Table 5 – Uranium and plutonium results for intercomparison seawater samples obtained on the 1 MV and on the 500kV compact AMS system at the CNA and ETH. The seawater samples come from two sampling stations in the Arctic Ocean (218 and 204) and different depths. 1 sigma errors are given.*

| Sample | Volume (L) | $^{236}$U (10$^6$ atoms kg$^{-1}$) | | $^{236}$U/$^{238}$U (x10$^{-9}$ atom ratio) | | $^{240}$Pu/$^{239}$Pu (atom ratio) | |
|---|---|---|---|---|---|---|---|
| | | CNA | ETH | CNA | ETH | CNA | ETH |
| 218_5m | 3.2 | 22 ± 2 | 21.9 ± 0.9 | 2.76 ± 0.13 | 2.81 ± 0.07 | 0.21 ± 0.05 | 0.27 ± 0.06 |
| 218_750m | 2.9 | 19 ± 1 | 20.1 ± 0.8 | 2.35 ± 0.14 | 2.22 ± 0.06 | 0.28 ± 0.04 | 0.21 ± 0.01 |
| 204_80m | 3.6 | 23 ± 1 | 23.9 ± 0.9 | 2.87 ± 0.15 | 2.92 ± 0.09 | 0.16 ± 0.04 | 0.16 ± 0.02 |
| 204_250m | 5.1 | 18 ± 1 | 19.6 ± 1.1 | 2.12 ± 0.10 | 2.40 ± 0.11 | 0.19 ± 0.04 | 0.19 ± 0.02 |

*Table 6 – Activity concentrations of $^{239}$Pu and $^{240}$Pu measured on the 1MV AMS system at CNA for seawater samples from Arctic Ocean.*

| Sample | $^{239}$Pu (µBq kg$^{-1}$) | $^{240}$Pu (µBq kg$^{-1}$) |
|---|---|---|
| 218_5m | 0.69 ± 0.07 | 0.54 ± 0.10 |
| 218_750m | 1.75 ± 0.13 | 1.82 ± 0.19 |
| 204_80m | 1.16 ± 0.12 | 0.71 ± 0.14 |
| 204_250m | 1.08 ± 0.09 | 0.74 ± 0.13 |

**Figures**

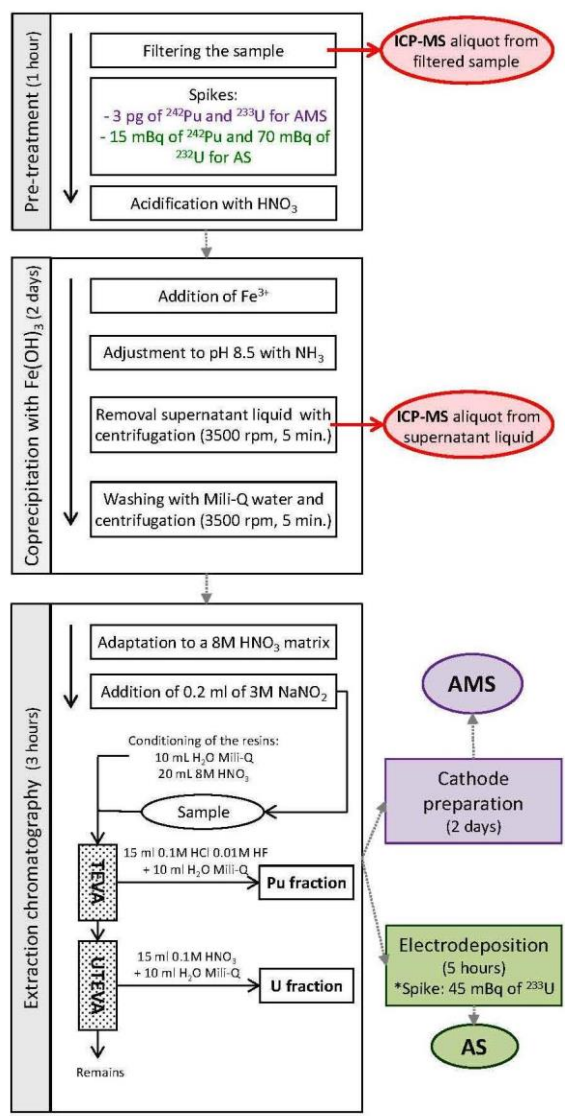

Fig. 1 – Flow-diagram of the radiochemical method proposed in this paper for plutonium and uranium measurement in seawater by AMS (purple). It is shown the adaptation of this method for AS measurements (green) and the aliquots extracted in order to study coprecipitation yield by ICP-MS (red).

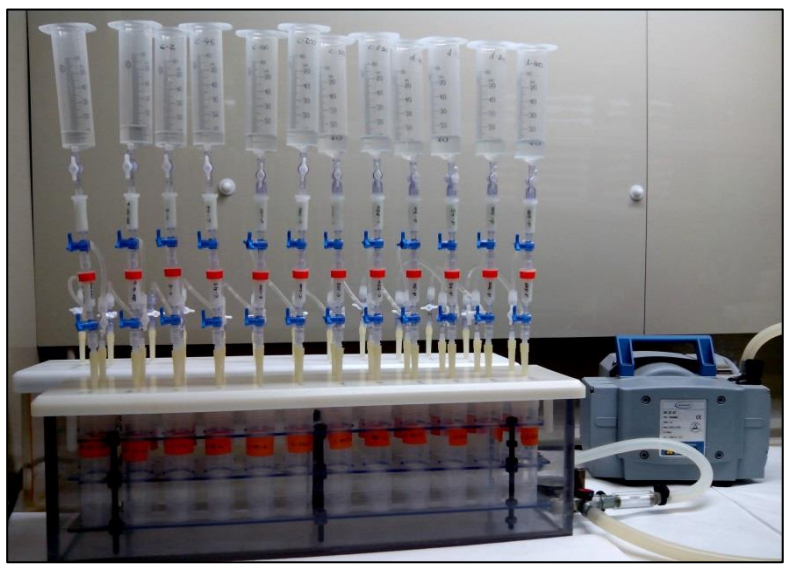

Fig. 2 – System performed for the extraction chromatography step: TEVA® and UTEVA® resin arranged in tandem in a vacuum box.

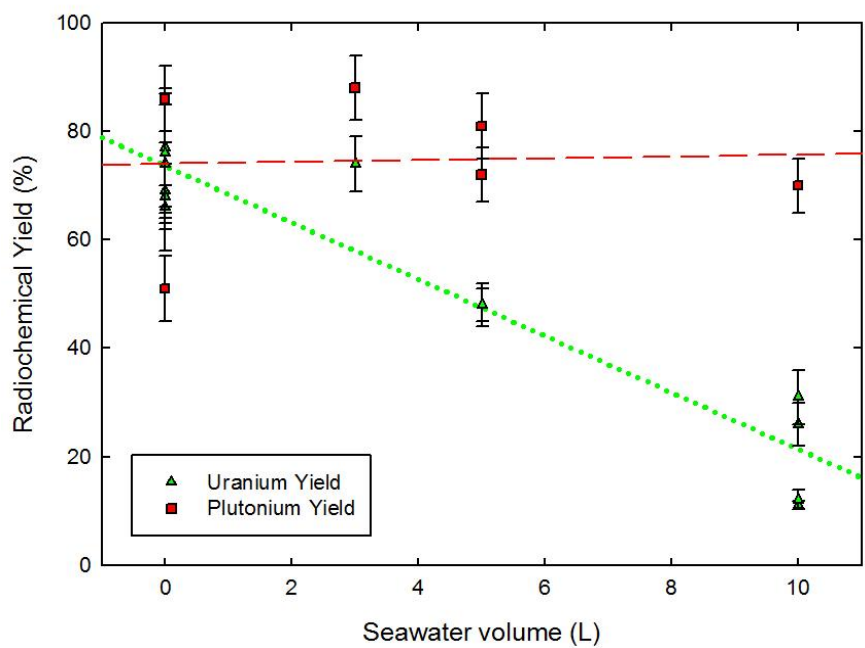

Fig. 3 – Radiochemical yields for uranium and plutonium isotopes obtained for different volumes of processed seawater.

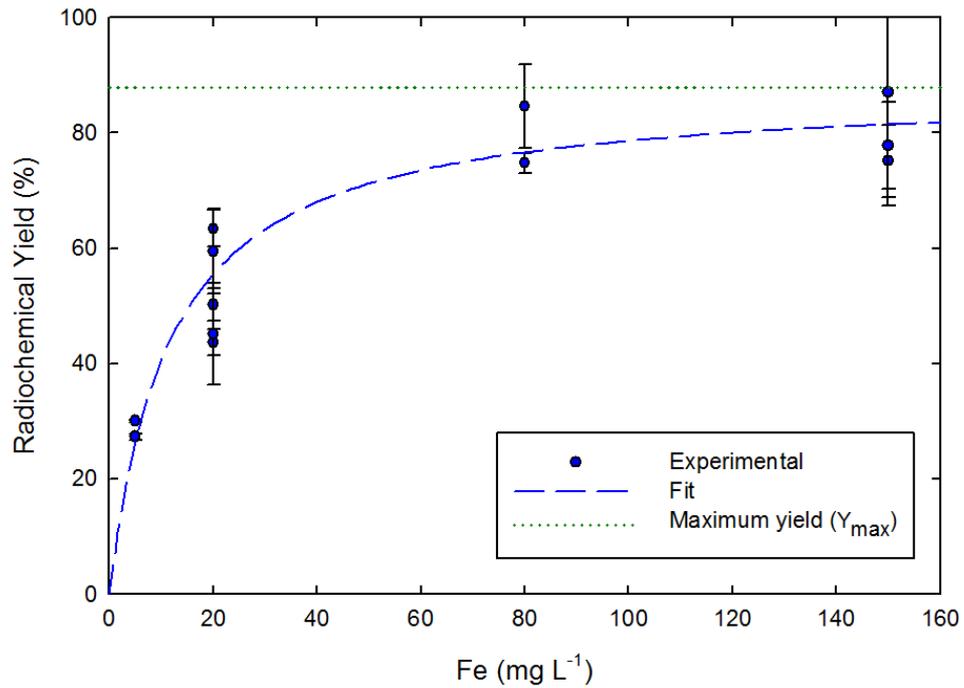

Fig. 4 – Uranium radiochemical yield of the coprecipitation process obtained by ICP-MS for different Fe(III) concentrations. All the experimental point came from one litre seawater samples.

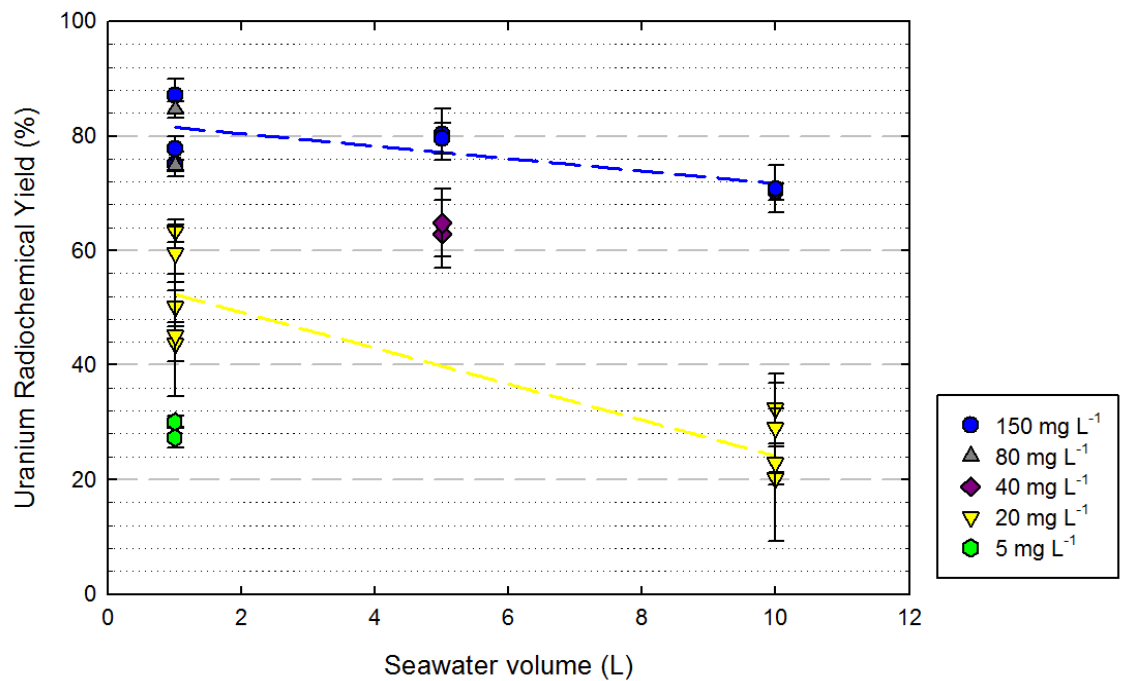

Fig. 5 – Uranium radiochemical yield of the coprecipitation process as a function of the total seawater volume for different Fe(III) concentrations. All these results have been obtained by ICP-MS.

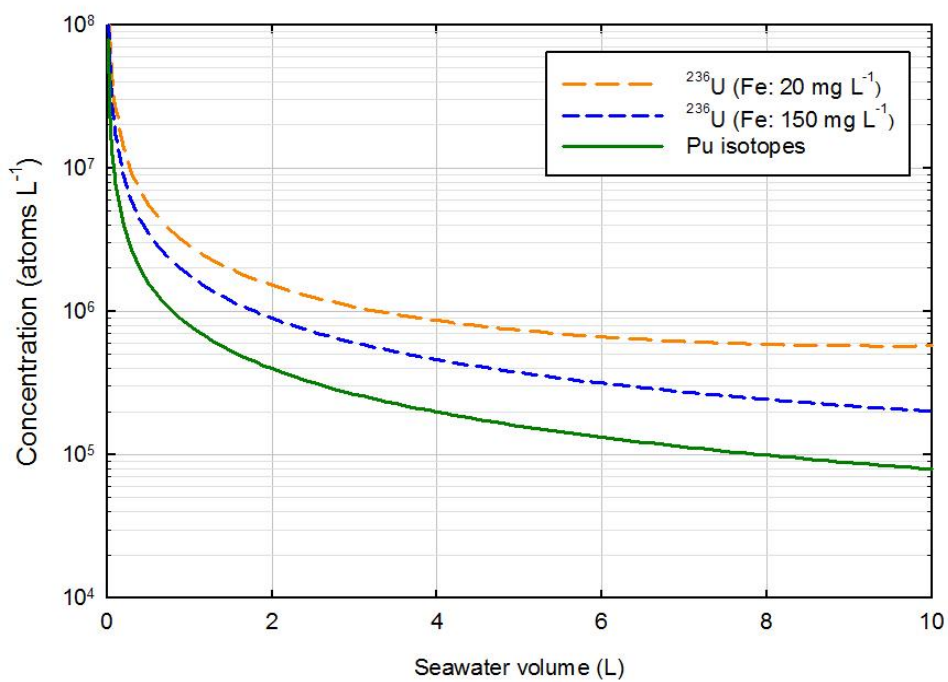

Fig. 6 – Minimum atomic concentrations for plutonium isotopes (green line) and for $^{236}$U (dashed lines), as a function of the seawater volume, that can be evaluated on the CNA AMS system with a statistical precision of 10%. For $^{236}$U measurements two cases for different Fe(III) concentrations in the coprecipitation step are plotted: 20 mg L$^{-1}$ (orange) and 150 mg L$^{-1}$ (blue).

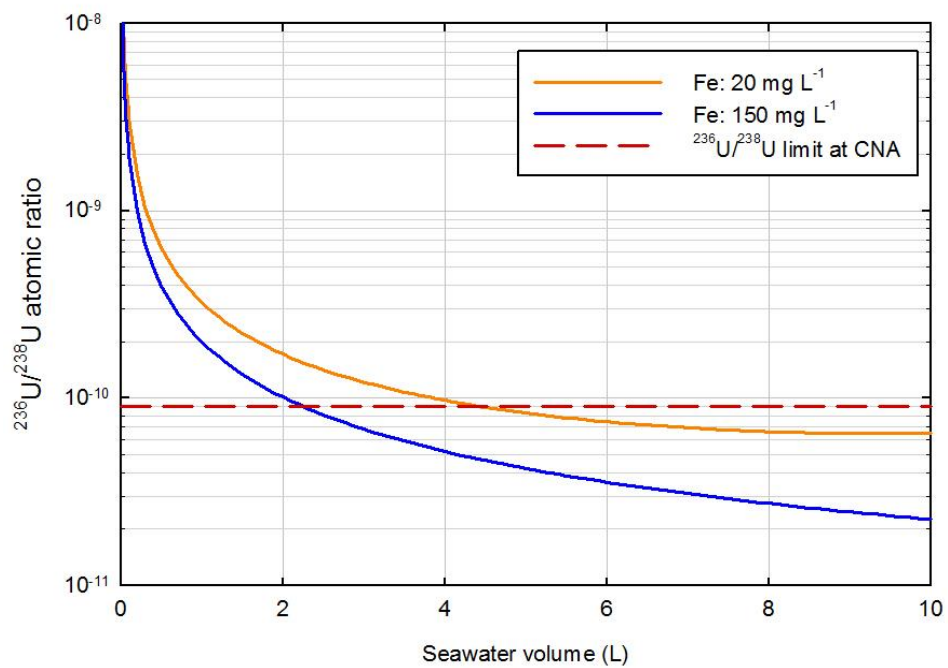

Fig. 7 – $^{236}$U/$^{238}$U atomic ratio as a function of the seawater volume that can be evaluated on the CNA AMS system with a statistical precision of 10%. Estimated recoveries by using 20 mg L$^{-1}$ (orange) and 150 mg L$^{-1}$ (blue) of Fe(III) in the coprecipitation step were considered. The red dashed line indicates the $^{236}$U/$^{238}$U limit of the AMS system at CNA.

## Data availability

The data that support the findings of this study are openly available at the following URL: https://github.com/AMS-CNA/Method-U-Pu-SW-2018/